\documentclass[
twocolumn,
]{openjournal}

\usepackage{graphicx}
\usepackage{dcolumn}
\usepackage{bm}
\usepackage{stfloats}

\usepackage{enumitem}
\usepackage{float}
\usepackage{wrapfig}
\usepackage{hyperref}
\hypersetup{
    unicode, 
    colorlinks=true,
    linkcolor=linkcolor,
    citecolor=linkcolor,
    filecolor=linkcolor,
    urlcolor=linkcolor,
}

\usepackage[table]{xcolor}
\definecolor{linkcolor}{rgb}{0.0,0.3,0.5}
\usepackage{tensind}
\tensordelimiter{?}
\DeclareGraphicsExtensions{.bmp,.png,.jpg,.pdf}
\usepackage{verbatim}
\usepackage[normalem]{ulem}
\usepackage{orcidlink}
\usepackage{soul}
\usepackage{afterpage}
\usepackage{placeins}

\begin{document}

\preprint{APS/123-QED}

\title{A Tale of Tails: \\ Star Formation and Stripping in Jellyfish Galaxies \\ in the Strong Lensing Cluster MACS J0138.0-2155}
\author{Catherine C. Gibson}
\affiliation{University of California, Santa Cruz, Santa Cruz, CA 95064, USA}
\affiliation{Santa Cruz Institute for Particle Physics, Santa Cruz, CA 95064, USA}
\author{Jackson H. O'Donnell}
\affiliation{University of California, Santa Cruz, Santa Cruz, CA 95064, USA}
\affiliation{Santa Cruz Institute for Particle Physics, Santa Cruz, CA 95064, USA}
\author{Tesla E. Jeltema}
\affiliation{University of California, Santa Cruz, Santa Cruz, CA 95064, USA}
\affiliation{Santa Cruz Institute for Particle Physics, Santa Cruz, CA 95064, USA}

\date{\today}
\begin{abstract}
We investigate the effects of ram-pressure stripping  on four galaxies within the massive, strong-lensing cluster MACS-J0138.0-2155 ($z=0.336$). Of these, three are classified as jellyfish galaxies, with significant elongated tails. Two of these jellyfish galaxies, J1 and J2, are in a late-stage of stripping and show post-starburst features within their disk regions with star formation only in the tails. Using VLT/MUSE integral field spectroscopic data, we spatially resolve the stellar and gas kinematics to examine extraplanar gas associated with ram-pressure stripping. We complement this analysis with optical and near-infrared imaging from the Hubble Space Telescope to visualize the galactic structure of each member. The jellyfish galaxies are all blue-shifted with respect to the cluster and show velocity gradients of a few hundred $\mathrm{kms}^{-1}$ across their tails. From the resolved gas kinematics, we derive H$\alpha$-based star formation rates; these are generally low reaching a maximum of approximately 0.49 $\mathrm{M_{\odot}\text{yr}^{-1}kpc^{-2}}$ in galaxy J3. We also report the kinematics for galaxy J4, which lies in the foreground of the cluster but close in projection to one of the lensed arcs.

\end{abstract}
\maketitle


\section{\label{sec:level1}Introduction\protect\\ }
It has long been known observationally that in the present epoch galaxies in dense environments like clusters of galaxies are different than those in under-dense environments, generally lacking recent star formation and having elliptical morphologies, so-called ``red and dead" galaxies \citep[e.g.][]{peng2010, mcpartland2016}.  While this difference can be partially explained by an earlier formation time for cluster galaxies, there are also environmental processes that can act to transform galaxies. For example, galaxy clusters provide the extreme environments necessary for ram-pressure stripping of cluster members, offering insight into the evolution and morphology of their member galaxies. Ram-pressure stripping (RPS) \citep{gunn1972} is caused by interactions between the intra-cluster medium (ICM) and the interstellar medium (ISM) of a galaxy that disrupt the hydrodynamic equilibrium, stripping away gas and dust from the outer layers of the galactic disk. 
This physical process is believed to generate the tentacle-like structure characteristic of jellyfish galaxies, a term coined to the best of our knowledge by \cite{bekki2009} and \cite{chung2009}.

Ram-pressure stripped tails are most commonly identified through radio continuum observations, H-$\alpha$ emission, and atomic hydrogen deficiencies \citep[e.g.][]{roberts2021, lee2022, scott2010}. Recent studies of galaxies within massive clusters at $z>0.2$ have revealed populations of jellyfish galaxies - cluster members exhibiting dramatic gas tails and morphological distortions - offering compelling evidence for active ram-pressure stripping in dense environments \citep[e.g.][]{ebeling2014, mcpartland2016, lee2022, moretti2022, owen2006, owers2012}. Evidence for RPS in cluster galaxies is derived from the presence of extraplanar ionized gas, a general lack of neutral hydrogen, and by contrasting the stellar and gas kinematics \citep{poggianti2025}.

Star formation rates above the star-forming main sequence point to triggering mechanisms beyond gravitational interactions like tidal stripping or mergers, potentially indicating that RPS is at play \citep[e.g.][]{poggianti2016, vulcani2018, vulcani2020, roberts2020}. RPS can drive rapid evolution in cluster galaxy members by locally enhancing and quenching star formation. This process is particularly prominent in jellyfish galaxies. Observationally, it has been shown that jellyfish galaxies exhibit extended tails of ionized gas that often can host in-situ star forming regions \citep[e.g.][]{fumagalli2014, cortese2007, george2018}.

Star formation quenching in jellyfish galaxies is linked to post-starburst galaxies because of their swift transition (in the last 1.5 Gyr) from active star formation to quiescence, likely caused by RPS. Post-starburst galaxies, often referred to as K+A galaxies, due to their spectral features resembling a combination of K-type stars and A-type stars, exhibit strong Balmer absorption lines and a lack of emission lines in their spectra \citep{dressler1982}. Studies of clusters using MUSE data have identified post-starburst signatures in galaxies undergoing RPS \citep{gullieuszik2017, werle2022}. 

While optical imaging like multi-band HST reveals the structural features of jellyfish galaxies, IFU data is effective in resolving the complex kinematics and physical conditions necessary for identifying star-forming regions, and the necessary conditions for accurate classification of jellyfish galaxies. Spatially resolved H$\alpha$ maps from IFU surveys like GASP \citep{poggianti2017, bellhouse2020, vulcani2018} reveal active star formation occurring directly within the stripped material of jellyfish galaxies. 

\begin{figure*}[hbtp]
    \centering  
    \includegraphics[width=\textwidth]{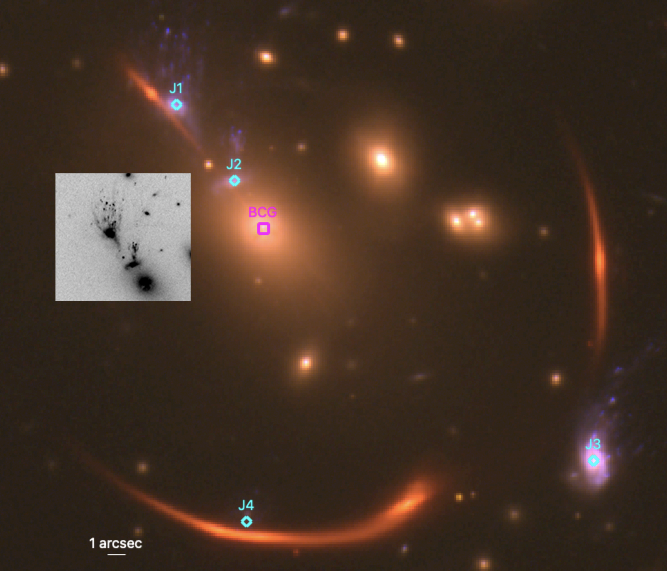}  
    \caption{RGB frame of MACS J0138.0-2155 cluster created from data retrieved from the Hubble Space Legacy Archive (Program ID: 14496, PI:Newman) with the red and green frames imaged using the WFC3 with the F160W and F105W filters, respectively, and the blue frame imaged with the ACS and the F555W filter. The RGB image was compiled using the image visualization and analysis tool SAO Image DS9 \citep{joye2017}. Jellyfish galaxies are marked with cyan diamonds and the BCG is marked with a magenta square. We overlay a snippet of the F555W frame with an inverted black and white color scheme to further emphasize the jellyfish structure observed in J1 and J2. Corresponding positions and redshifts for each galaxy can be found in Table~\ref{tab:systems}.}  
    \label{fig:HST}  
\end{figure*}

\begin{table*}[t!]
\caption{\label{tab:systems} The right ascension, declination, redshift, and stellar velocity dispersion of each jellyfish galaxy as well as the brightest central galaxy within the MACS J0138.0-2155 galaxy cluster labeled as shown in Fig.~\ref{fig:HST}. The regions used to find the corresponding redshifts encompass all Voronoi bins in Fig.~\ref{fig:Bins} for each galaxy listed here. RA and DEC are listed in sexagesimal coordinates.}
\begin{ruledtabular}
\begin{tabular}{ccccc}
 Label & RA & DEC & $z$ & $\sigma_{stellar}$
\\ \hline
 BCG& 1:38:03.7641 & -21:55:31.912 &0.3378 & 368 $\pm 2$\\
 J1&1:38:04.1203
 &-21:55:24.746&0.3323& 38 $\pm 3$\footnotemark[1] \\
 J2&1:38:03.8846&-21:55:29.120
 &0.3317 & 134 $\pm 3$\footnotemark[1]\\
 J3&1:38:02.3992&-21:55:45.190&0.3336& 98 $\pm 2$\\
 J4&1:38:03.8327&-21:55:48.705&0.3087& 35 $\pm 15$
 \footnotetext[1]{The stellar velocity dispersion listed for J1 and J2 are for the heads of the galaxy only and do not encompass the tail regions (these regions are the sum of the green bins in Fig.~\ref{fig:Bins} that do not overlap with the magenta tail bins for J1 and J2). The redshifts listed for J1 and J2 are for the entire galaxies (head+tail). Redshifts for the heads of J1 and J2 can be found in Sec.~\ref{sec:J1_results} and Sec.~\ref{sec:J2_results}.}
\end{tabular}
\end{ruledtabular}
\end{table*}

In this paper, we investigate the impact of ram-pressure stripping (RPS) on star formation in four member galaxies of the strong-lensing cluster MACS J0138.0–2155. To the best of our knowledge, three of these galaxies were first informally labeled P1, P2, and P3 by \cite{rodney2021}, who incorporated them in the lens modeling. We identify three of the four galaxies as jellyfish galaxy candidates, two of which exhibit characteristics of post-starburst galaxies. While the evolutionary stages of RPS are not precisely defined, we apply the morphological classification scheme proposed by \cite{poggianti2025} to interpret their stripping stages. Using integral field spectroscopy, we analyze ionized gas kinematics traced by emission lines and spatially resolve regions of active star formation. In section~\ref{sec:data} we will describe the data and sample used in this analysis, as well as background on the MACS J0138.0-2155 cluster, in section~\ref{sec:analysis} we will describe the methodology used to fit the spectra and derive star formation rates, and in section~\ref{sec:Results} and section~\ref{sec:discussion} we will interpret the spatially resolved kinematics and fits for each galaxy. 

\section{Data and Sample} 
\label{sec:data}
\subsection{Jellyfish Galaxies in MACS J0138.0-2155}

We identify four potential jellyfish galaxy candidates in the host galaxy cluster MACS J0138.0-2155 (z=0.336), first identified in the MAssive Cluster Survey (MACS; \cite{ebeling2001}). This galaxy cluster hosts two strongly-lensed supernova in the high-redshift source galaxy - SN Requiem \citep{rodney2021} and SN Encore \citep{pierel2024} - making it a prime target for repeated observations. Previous papers have presented kinematic modeling of the cluster galaxy members and the Faber-Jackson relation \citep{faber1976} for the cluster \citep{flowers2024,granata2024,acebron2025}. For the analysis in this paper, we label the galaxies of interest within this cluster as J1, J2, J3, and J4 as shown in Fig.~\ref{fig:HST}. 
\newcommand{\sharedfootnotemark}{\footnotemark[1]}
\newcommand{\sharedfootnotetext}{\footnotetext[1]{The stellar velocity dispersion reported for J1 and J2 are for the head regions only and do not include the tails.}}

\begin{figure}[h!]  
    \centering  
    \includegraphics[width=\columnwidth]{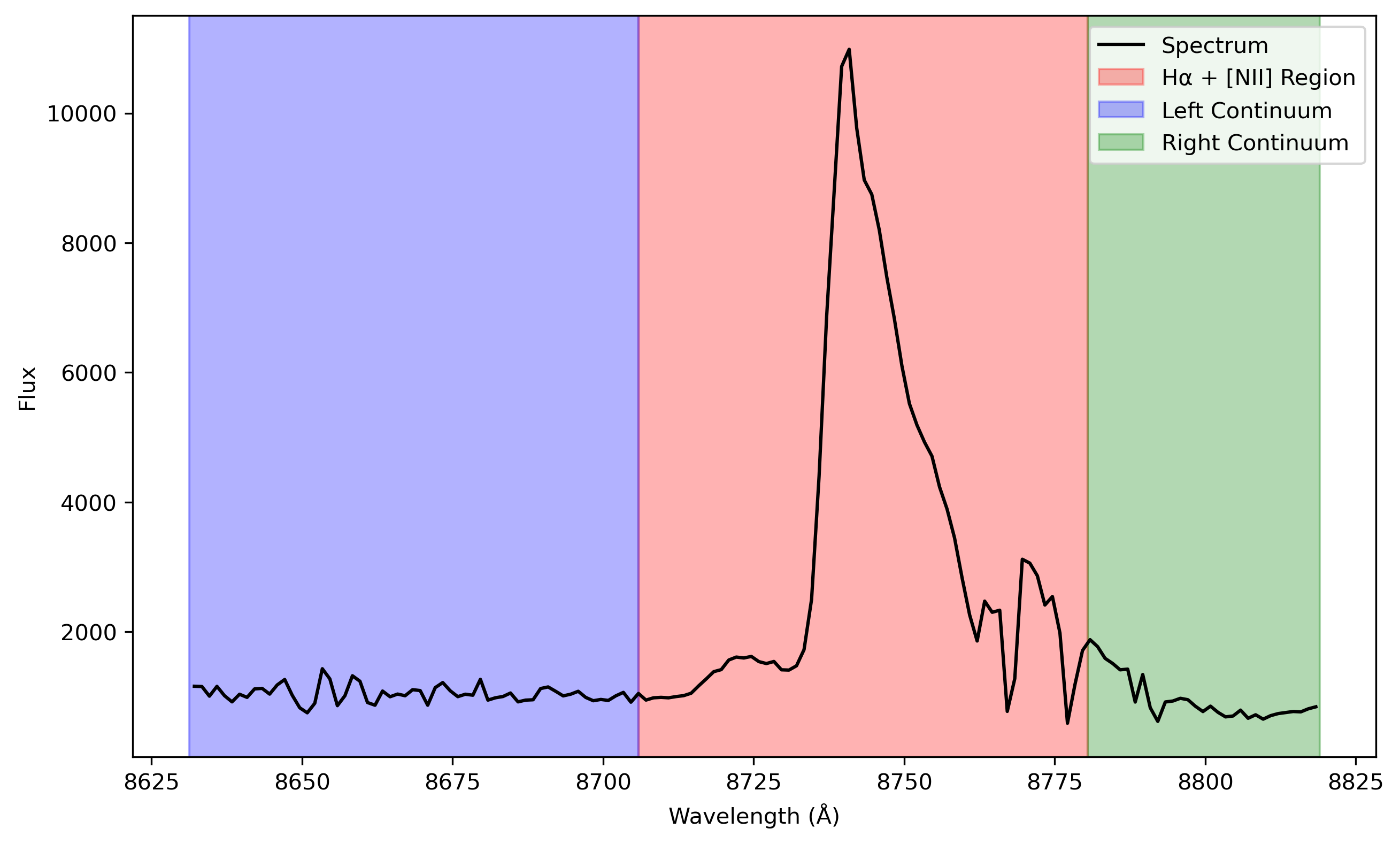}  
    \caption{Example of the H$\alpha + $[N II] signal with respect to the regions selected for continuum in the tail of the J1 galaxy. The left continuum region spans the rest frame wavelength range from 6478-6534~\AA\ and the right continuum region spans the rest frame wavelength range from 6590-6618~\AA. The S/N calculated for the region of the H$\alpha + $[N II] is scaled for disproportionate selected continuum regions on either side of the doublet. The disproportionality of the regions stems from noise in the data that was excluded. }  
    \label{fig:Signal}  
\end{figure}

\subsection{MUSE Spectroscopy}
 MACS J0138.0-2155 has been the target of two observation blocks with MUSE, the Multi-Unit Spectrographic Explorer, on the ESO Very Large Telescope UT4 \citep{bacon2010}. The IFU data cube analyzed in this study was obtained from the ESO archive and combines two sets of MUSE observations: a 49-minute exposure from 2019 (Program ID: 0130.A-0777, PI: A. Edge) and a 2.9-hour exposure from 2023 (Program ID: 110.23PS, PI: S.H. Suyu). The MUSE instrument covers a 1x1 arcmin$\mathrm{^2}$ field of view at 0.2x0.2 arcsecond pixel sampling \citep{bacon2010}. Details of the data reduction process, including the use of the MUSE data reduction pipeline \citep{weilbacher2020}, are described in \cite{granataCosmologySupernovaEncore2024a}. In particular, they note that the wavelength range from, 5800~\AA\ to 5965~\AA\ was masked in the latter dataset due to contamination from the GLAO laser guide system, part of MUSE's adaptive optics.

We also ran the Zurich Atmosphere Purge, \texttt{ZAP}, as an enhancement to the processed IFU spectroscopic data cube to improve sky subtraction residuals \citep{soto2016}.

\begin{figure}[h!]  
    \centering  
    \includegraphics[width=0.89\columnwidth]{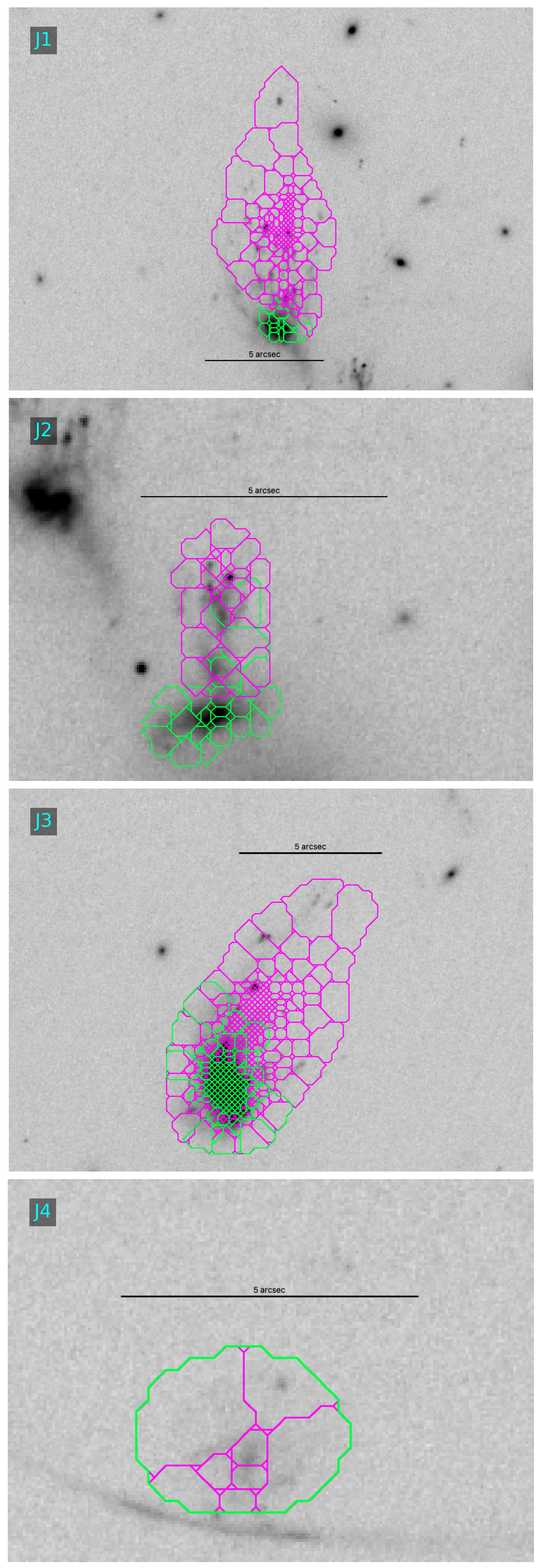}  
    \caption{From top to bottom, the inverted HST F555W filtered frame of the galaxies J1-4 with the stellar component Vorbins overlaid in green and the gas component Vorbins overlaid in magenta. The black scale bar indicates 5 arcseconds.}  
    \label{fig:Bins}  
\end{figure}

\subsection{Jellyfish Structure and Redshifts}
Hubble Space Telescope near-infrared and visible imaging of this galaxy cluster in the F555W, F105W, and F160W filters with each galaxy marked for analysis can be found in Fig.~\ref{fig:HST}. Their corresponding positions, redshifts, and stellar dispersions are listed in Table~\ref{tab:systems}. We estimate the projected distances of each galaxy to the BCG for J1, J2, J3, and J4 as approximately 42, 15, 111, and 81 kpc, respectively.
To find the preliminary redshifts of each marked galaxy, we utilize the \texttt{Marz} spectral analysis tool to match each galaxy's spectra against stellar and galactic templates \citep{hinton2016}.  

We make the distinction between a well-defined disk and gas-stripped tail in galaxies J1 and J2 from inspection of their visual structure in the HST imaging. We will refer to these regions as the ``head" and ``tail" for J1 and J2 throughout this paper. The tails of J1 and J2 were estimated to be $\sim$21 kpc and $\sim$16 kpc in length, respectively. The heads were estimated to be $\sim$5 kpc in width for J1, and $\sim$4 kpc in width for J2. We list redshifts found using \texttt{Marz} \citep{hinton2016} for each galaxy in Table~\ref{tab:systems}, but also find individual redshifts for the heads and tails of J1 and J2. As later described in Section~\ref{sec:spec}, for galaxies J1 and J2, we end up using the head redshifts for the fitting of the continuum-binned Voronoi regions, and the tail redshifts for the fitting of the gas emission-binned Voronoi regions.

\section{Analysis}
\label{sec:analysis}
\subsection{Binning}
In order to accumulate sufficient statistics for spectral analysis, we bin the MUSE data cube into 2D spatial regions based on signal-to-noise. Specifically, we use the Python package \texttt{VorBin}, an adaptive spatial binning method \citep{cappellari2003}. We utilize two separate binning schemes, yielding two sets of spatial bins per galaxy to minimize uncertainties on the gas and stellar kinematics. The emission binning scheme allows us to extract star formation rates in each bin, and our continuum binning allows us to accurately represent absorption features in each bin. 

To study the stellar component in each galaxy, we perform spatial binning in each galaxy based on the S/N of the continuum in the observed wavelength range from 6000-6150~\AA\, adopting a target S/N of 20. We apply this binning scheme to each galaxy, and remove some bins from our stellar kinematics results following the cuts described in Section~\ref{sec:spec}. The remaining continuum bins used for our stellar fits, are shown in green, overlaid on the HST imaging of each galaxy in Fig.~\ref{fig:Bins}. 

We apply binning based on the H$\alpha + $[N II] doublet to J3, J4, and the tails of J1 and J2 to study the gas component and estimate star formation rates in each galaxy. The heads of J1 and J2 are not included in the emission binning scheme because they lack any strong emission features. To bin by the H$\alpha$ emission, we first estimate the line fluxes with the following continuum subtraction procedure. For galaxies that exhibit strong H$\alpha + $[N II] emission, we define the signal region as the rest-frame wavelength range associated with the doublet, from 6534-6590~\AA\!, relative to the stellar continuum. Fig.~\ref{fig:Signal} illustrates the S/N calculation used for Voronoi binning in the tail of the J1 galaxy. The continuum used spans 6478-6534~\AA\ on the left side of the H$\alpha$ signal and 6590-6618~\AA\ on the right side. The S/N is then scaled for the disproportionate continuum selection on either side of the signal. The chosen signal and continuum regions remain consistent across these galaxies in the rest-frame wavelength range, and we utilize a target S/N of 25 in our Voronoi binning scheme around the H$\alpha + $[N II] emission feature. The resulting Vorbins used for fitting the gas component are shown in Fig.~\ref{fig:Bins}, overlaid on each galaxy in magenta. 

\subsection{Spectral Fitting using pPXF} 
To fit the spectra of each Voronoi bin, we used the penalized pixel fitting (\texttt{pPXF}) method \citep{cappellari2004, cappellari2017, cappellari2023}. This method fits an observed spectrum using a linear combination of gas and stellar templates convolved with a line-of-sight velocity distribution (LOSVD), allowing us to extract stellar and gas kinematics. To ensure accurate velocity measurements, we provided an initial redshift from \texttt{Marz}, which \texttt{pPXF} refined by computing velocity offsets for each Voronoi bin. We supplied \texttt{pPXF} with stellar templates from XSL DR3, the stellar spectral library from the X-Shooter spectrograph on the Very Large Telescope \citep{vernet2011, verro2022}. For all stellar templates, we set the signal-to-noise threshold per spaxel to 3. Gas emission lines were modeled using Gaussian templates, each with its own LOSVD. We also have the option to define masked regions as well as multiplicative or additive polynomials to fit the continuum rather than drawing from a stellar spectral library. For our analysis, we considered templates in the wavelength range 3500-6900~\AA\ with a normalization range of 5400-5500~\AA. In some cases, we adjust the start to 3600~\AA\ and the normalization range from 5300-5400~\AA, depending on the redshift of the galaxy and the noise features present in the continuum. We mask out the portion of the spectrum that was removed due to contamination from the GLAO laser star guide system in the observed wavelength range from 5800-5965~\AA\ in all galaxies. We set the additive Legendre polynomial used to correct the template continuum for any unfitted background flux to degree 6 in all fits. Additionally, we select varying wavelength ranges to mask out, depending on noise features in the continuum per galaxy.

\subsubsection{Emission Line Fitting and Absorption Features}
\label{sec:spec}
In this analysis, we use two kinematic components to fit the observed spectra. The stellar kinematic component models the continuum and captures the absorption line features while the gas kinematic component is used to capture the gas emission lines. We focus on fitting the H$\alpha$ (6563~\AA), H$\beta$ (4861~\AA), and [NII] (6548~\AA\ \& 6584~\AA) emission lines.

To avoid overfitting emission lines, we fix the stellar continuum with an optimal stellar template derived from fitting the integrated spectrum of all spaxels in each galaxy with a best-fit linear combination of stellar templates, allowing only its amplitude, kinematics, and emission line properties to vary per individual Voronoi bin. The gas emission lines are modeled with their own Gaussian templates with independent flux amplitudes; however, all emission lines in the same Vorbin are tied kinematically. 

This fitting method is applied across all Voronoi bins, in both binning schemes. We report the stellar kinematics per each galaxy using the stellar component and the continuum binned scheme, and we report the gas kinematics using the gas component and the H$\alpha+$[NII] Vorbins. We do not report results from the gas component for the continuum-binning scheme, because these bins do not necessarily have significant emission-line flux.

In the continuum-binning scheme, some Voronoi bins are dominated by strong emission and weak or absent stellar absorption, yielding unreliable fits and therefore unreliable stellar kinematics. To mitigate large uncertainties in these cases, and ensure that the reported stellar kinematics are robust, we exclude all Voronoi bins with $\chi^2 > 1.5$ and those with fractional uncertainties on the stellar velocity dispersion exceeding $40\%$. This ends up eliminating most of the tail bins in J1 and J2, and maintaining most of the head bins. Because of this, we utilize the redshift of the heads of J1 and J2 as the initial input redshift to \texttt{pPXF} rather than the redshift of the entire galaxies (head+tail). This is mainly to have a good starting scale for fitting the stellar velocity offset values. 

\begin{figure*}[!ht]
    \centering
    \includegraphics[width=\textwidth]{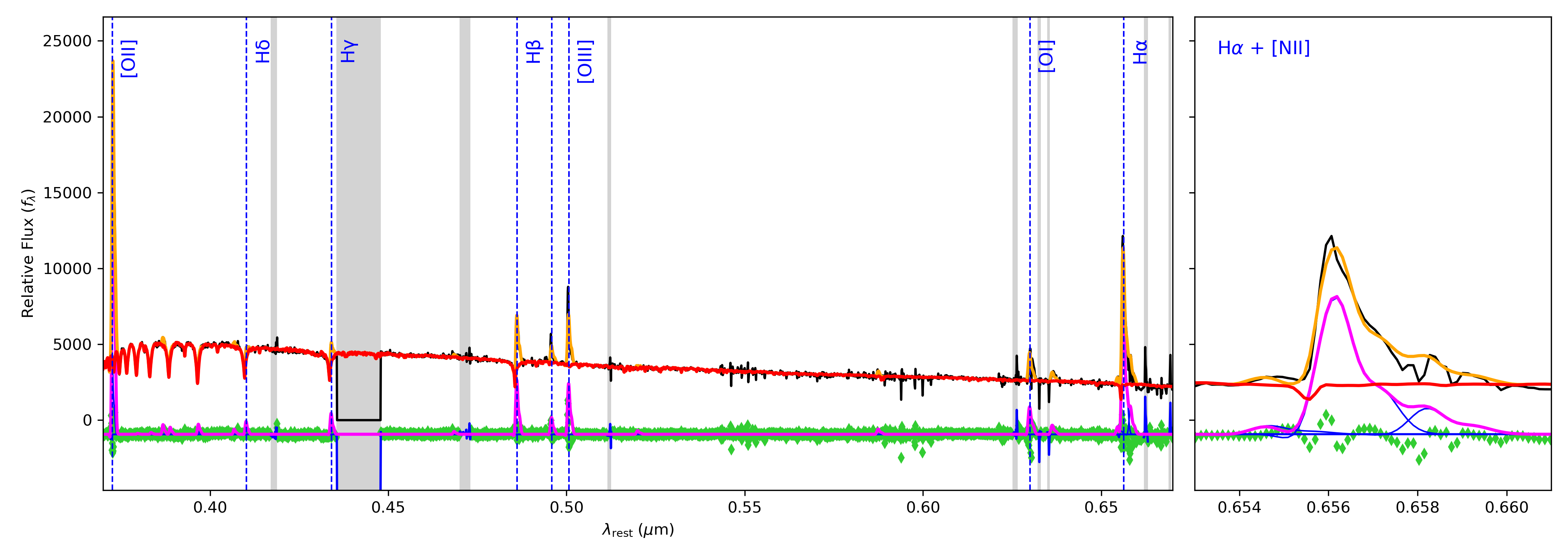}  
    \caption{This figure shows the combined \texttt{pPXF} fit of all of the Voronoi bins of the J1 galaxy, including the head and tail. The leftmost plot shows the relative flux of the observed spectrum in the rest frame wavelength range from 3500-6900~\AA. The zoomed-in fit of the H$\alpha +$N[II] emission lines is shown on the right-hand side. Black indicates the observed spectra, red is the fit of the stellar continuum, orange is the emission line template, green is the fit residual, and pink is the combined fit of the stellar and gas templates. The stellar continuum is fitted with an optimal template derived from the XSL DR3. The grey regions are the wavelength ranges masked from our fit, with their corresponding residuals in blue. The spectral features of interest are labeled and marked in blue, dashed lines on the plot. The combined \texttt{pPXF} fit of J1 does not vary based on the binning scheme, as all bins are included.}
    \label{fig:J1_spectra}
\end{figure*}

\begin{figure*}
    \centering
    \includegraphics[width=\textwidth]{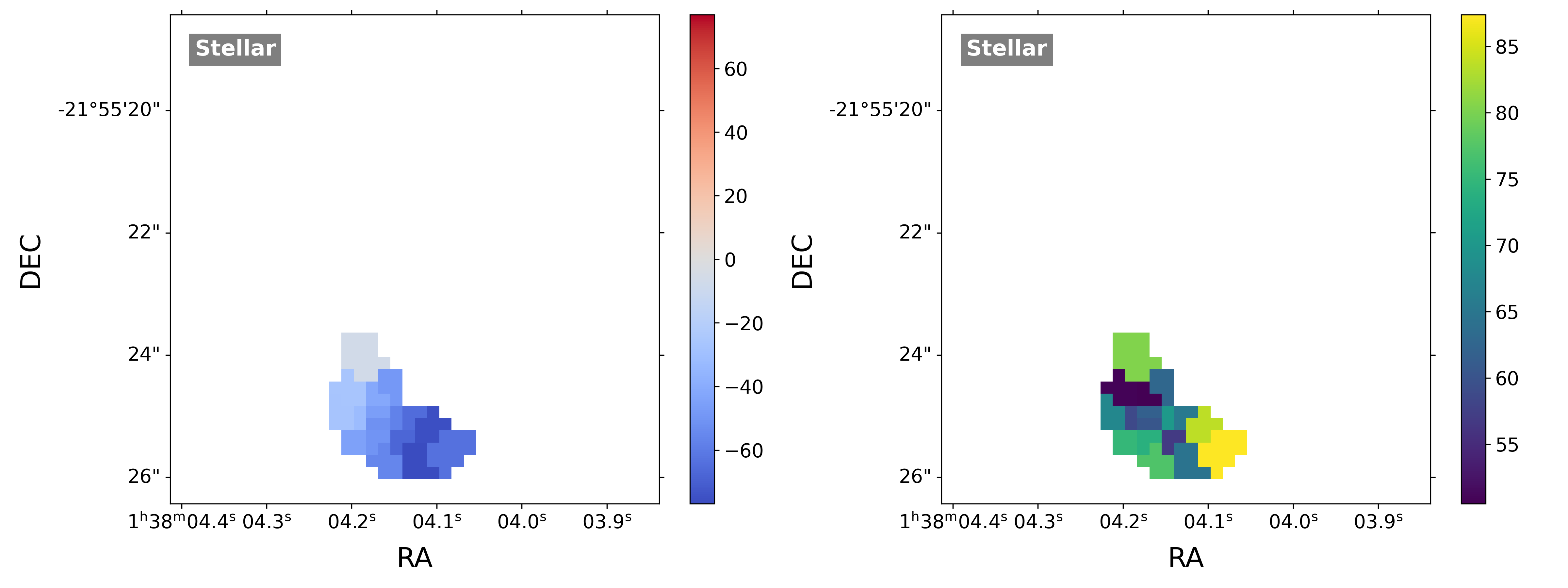}  
    \caption{The line-of-sight stellar velocity offset (left) and dispersion (right) in $\mathrm{km\, s^{-1}} $ for the continuum-binned Voronoi bins in the J1 galaxy after relevant quality cuts. The reported velocity offsets here are with respect to the redshift of the head ($z=0.3310$). }
    \label{fig:J1_head_kin}
\end{figure*}

\subsection{SFRs}
\label{sec:SFR}
To derive the star formation rates, we use the luminosity of the H$\alpha$ emission line corrected for stellar absorption, foreground extinction, and internal dust extinction. Stellar absorption corrections are already accounted for by \texttt{pPXF} in the fitting process. To correct for dust extinction, we utilize the Balmer decrement, the absorption-corrected flux ratio of the H$\alpha$ and H$\beta$ lines. We adopt an intrinsic H$\alpha$ to H$\beta$ ratio of 2.86 and adopt the \cite{cardelli1989} extinction law. The extinction magnitude of the flux of H$\alpha$ is given by 
\begin{equation}
A_{H_{\alpha}} = k_{H_{\alpha}} \times E(B-V),
\end{equation}
where $k_{H_\alpha}$ is the extinction coefficient which we will adopt as 2.53 from \cite{cardelli1989} and $E(B-V)$ is the color excess which can be found from the following equation, 
\begin{equation}
E(B-V) = 2.33 \times \log \bigg[\frac{S_{H_\alpha}/S_{H_{\beta}}}{2.86}\bigg]
\end{equation}
where $S_{H_\alpha}/S_{H_\beta}$ is the absorption-corrected flux ratio. To correct for foreground Galactic extinction, we used the NASA/IPAC Extragalactic Database \href{https://ned.ipac.caltech.edu/extinction_calculator}{NED Extinction Calculator}, which applies recalibrated dust maps from \cite{schlafly2011} based on the original work of \cite{schlegel1998}. We select the SDSS z-filter for 0.8923 $\mu$m with the galactic extinction of 0.020 mag. Using the absorption, dust, and foreground corrected H$\alpha$ flux, we can now find the star formation rates using the following relation given by \cite{kennicutt1998}, 
\begin{equation}
\text{SFR}_{H_{\alpha}} = 4.6 \times 10^{-42}L_{H_\alpha},
\end{equation}
where $\text{SFR}_{H_{\alpha}}$ is in units of $M_{\odot}\text{yr}^{-1}$, and $L_{H_{\alpha}}$ is the corrected H$\alpha$ luminosity. We report all SFRs in units of $\mathrm{M_{\odot}\text{yr}^{-1}kpc^{-2}}$, such that the calculated star formation rates are normalized by the number of pixels per Voronoi bin and pixel area. We adopt the IMF given by \cite{chabrier2003} for the SFR calculations and assume a flat $\Lambda$CDM cosmology of $H_{0}=70\mathrm{km\, s^{-1}\, Mpc^{-1}}$, $\Omega_{M} = 0.3$ and $T_{\text{CMB}} = 2.725 \mathrm{K}$ for conversion from H$\alpha$ flux to luminosity, and pixel area to $\mathrm{kpc^{2}}$. 

\begin{figure*}[!ht]
    \centering
    \includegraphics[width=\textwidth]{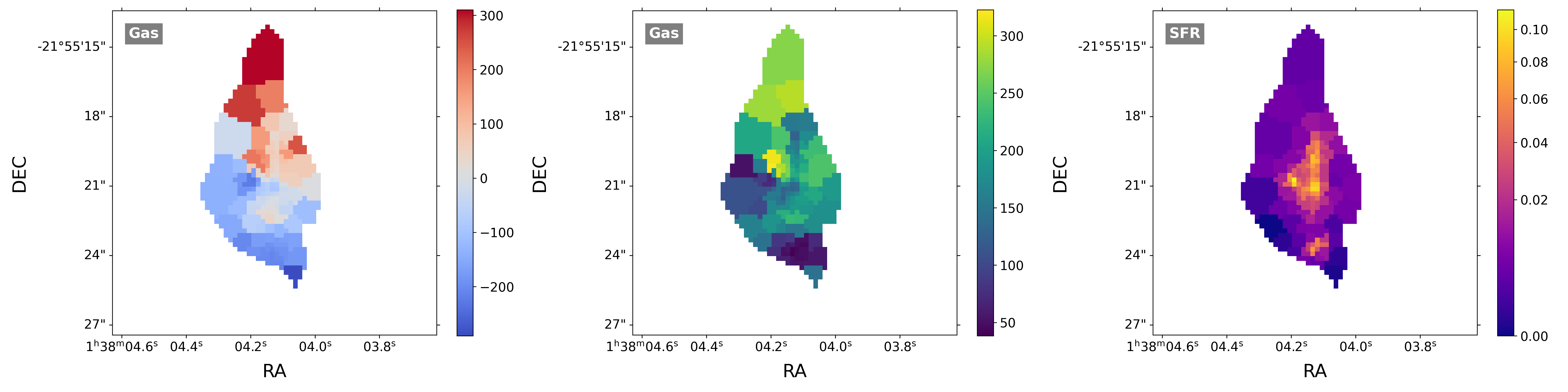}
    \caption{The leftmost panels display the spatially resolved gas velocity offset and gas velocity dispersion (from left to right) in $\mathrm{km s^{-1}}$ of the observed spectrum of all Voronoi bins in the J1 galaxy tail. The velocity offset is measured with respect to the redshift of the tail (z=0.3324). The rightmost plot displays the star formation rates in each bin as calculated in Section~\ref{sec:SFR} in units of $\mathrm{M_{\odot}\text{yr}^{-1}kpc^{-2}}$, with the color scale shown in square-root stretch. }
    \label{fig:J1_tail_kinematics}
\end{figure*}

\begin{figure*}[htbp]
    \centering
    \includegraphics[width=\textwidth]{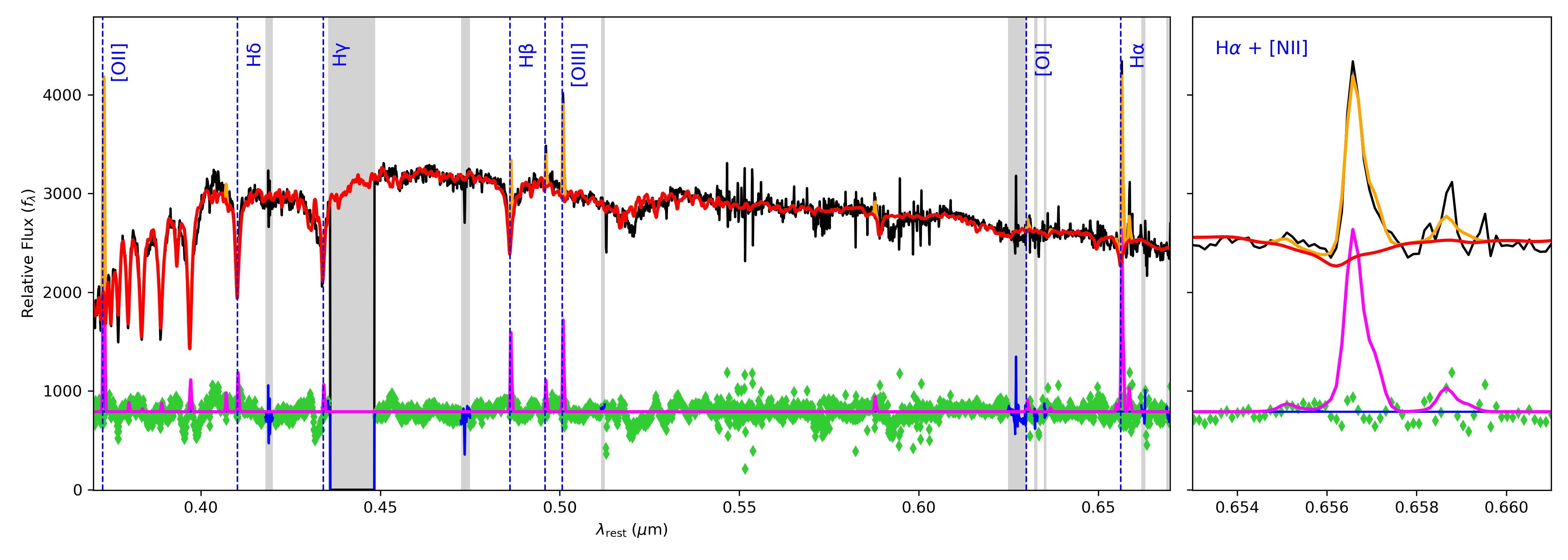}
    \caption{The combined \texttt{pPXF} fit of all Voronoi bins in J2 irrespective of binning scheme choice. The zoomed-in fit of the H$\alpha+$[NII] emission doublet is shown in the rightmost panel. The color scheme follows that of Fig.~\ref{fig:J1_spectra}.}
    \label{fig:J2_spectra}
\end{figure*}

\begin{figure*}[htbp]
    \centering
    \includegraphics[width=\textwidth]{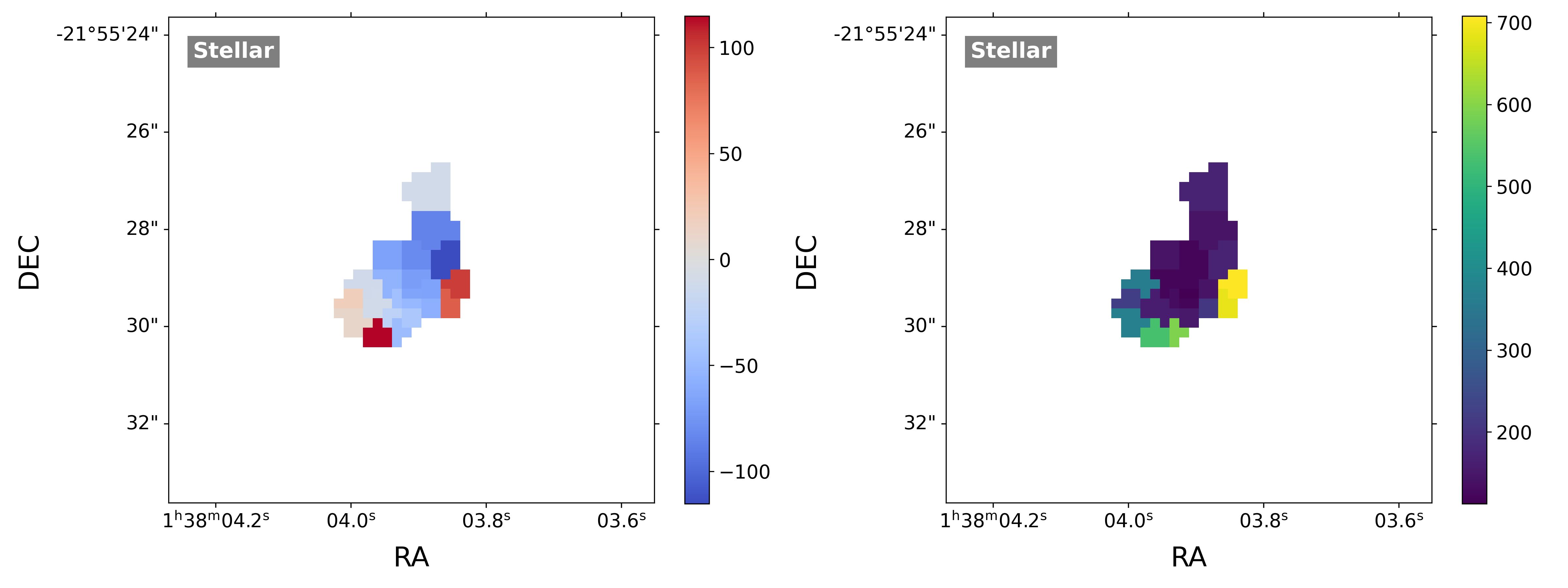}  
    \caption{The line-of-sight stellar velocity offset and dispersion in $\mathrm{km {s}^{-1}}$ from left to right of all continuum binned Voronoi regions in the J2 galaxy after quality cuts have been applied. The velocity offset is measured with respect to the redshift of the head region of J2 (z=0.3308).}
    \label{fig:J2_head_kin}
\end{figure*}

\section{Results}
\label{sec:Results}
In this section, we summarize the results for each individual jellyfish galaxy. Average redshifts and stellar velocity dispersions for each galaxy are given in Table~\ref{tab:systems}. All four galaxies are blue shifted relative to the BCG. J4 appears to be in the foreground of the cluster, while J1-3 are consistent with infalling into the cluster. As discussed below, we see additional velocity gradients across the tails of these three jellyfish galaxies. As previously mentioned, the continuum-binned Vorbins are utilized for fitting the stellar kinematics in each galaxy, and the H$\alpha$-binned Vorbins are used for fitting the gas kinematics and calculating the star formation rates. 

\subsection{J1}
\label{sec:J1_results}

The J1 galaxy exhibits a characteristic jellyfish morphology, with a distinct head and an ionized, stripped tail, clearly visible in the HST imaging. J1 is approximately 42 kpc from the BCG. We measure the width of the head of J1 to be approximately 5 kpc and the tail to be approximately 21 kpc in length. We use two different redshifts in the fitting for J1, the redshift of the head (z=0.3310) for fitting our continuum-binned Voronoi regions, and the redshift of the tail (z=0.3324) for fitting our H$\alpha$-binned Voronoi regions. We report the average redshift of the entire J1 region including the head and tail as z=0.3323, which we did not use in our analysis fitting, but is listed in Table~\ref{tab:systems}. We separate the redshifts by head and tail for our two binning schemes for J1, because stellar absorption is almost entirely limited to the head region, and gas emission is limited to the tail and the head and tail are significantly offset in velocity. These initial redshifts are fed to \texttt{pPXF} which are then utilized to determine the relative velocity offsets within each Voronoi bin in the fitting process.

The combined \texttt{pPXF} fit of the entire J1 galaxy, including the head and tail, is shown in Fig.~\ref{fig:J1_spectra}. For the full fit of J1, we do not apply any quality cuts or selection criteria, to most accurately represent the fit of the entire observed spectrum. The Balmer absorption present in the spectrum can be traced to the spectral features present in the head, while the strong H$\alpha$ emission can be traced to the spectral features in the tail of J1. We expand on the features present in the head in Section~\ref{sec:discussion}. 

To study the stellar kinematics in J1, we utilize the stellar component of the \texttt{pPXF} fit of each Voronoi region created using the continuum binning scheme. 
The stellar continuum is fit using an optimal stellar template derived from XSL DR3 fit to the integrated spectrum of all spaxels used for binning in J1.  
We present the stellar kinematics for the J1 bins that meet our selection criteria as described in Section~\ref{sec:spec} in Fig.~\ref{fig:J1_head_kin}. The surviving bins are clearly limited to the head region and appear to be relatively undisturbed. 
The velocity offset is on the order of approximately $-50\ \mathrm{km s^{-1}}$ relative to the initial redshift with a slight gradient across the head potentially indicating rotation. 

The gas kinematics for J1 are fit using the Voronoi regions created using the signal of the H$\alpha +$N[II] emission doublet for binning. In this binning scheme, the head region of J1 did not have enough signal in the H$\alpha +$N[II] to be included, and we removed this region from our gas fit. The Voronoi bins used for measuring the gas kinematics in J1 are shown in magenta in Fig.~\ref{fig:Bins}. We present the gas kinematics for each Voronoi bin as well as the SFRs for the J1 tail in Fig.~\ref{fig:J1_tail_kinematics}. We observe a clear velocity gradient across the tail in the leftmost panel of Fig.~\ref{fig:J1_tail_kinematics}, indicative of stripping. The mean line-of-sight velocity of the tail is offset by more than $300\ \mathrm{km s^{-1}}$ with respect to the head, and there is a gradient of more than $400\ \mathrm{km s^{-1}}$ across the tail with the side of the tail closest to the head also being closer in peculiar velocity. We also find a gradient present in the velocity dispersion across the tail with the velocity dispersion generally rising as distance from the head increases. The rightmost panel in Fig.~\ref{fig:J1_tail_kinematics} shows the star formation rate with a prominent star-forming knot present near the center of the tail, and a smaller knot towards the base of the tail. The star formation rates reach approximately 0.11 $\mathrm{M_{\odot}\text{yr}^{-1}kpc^{-2}}$ and taper off from the central knot.  The statistical uncertainty on the SFR values are typically $\sim$4\% of the reported SFR values in $\mathrm{M_{\odot}\text{yr}^{-1}kpc^{-2}}$. In the regions surrounding the smaller star forming knot near the base of the tail, the uncertainty on the SFR reaches $\sim$10-15\% of the reported SFR value. We note that the velocity dispersion is about 150 $\mathrm{km s^{-1}}$ in the spatial region where star formation is strongest.

\begin{figure*}[t]
    \centering
    \includegraphics[width=\textwidth]{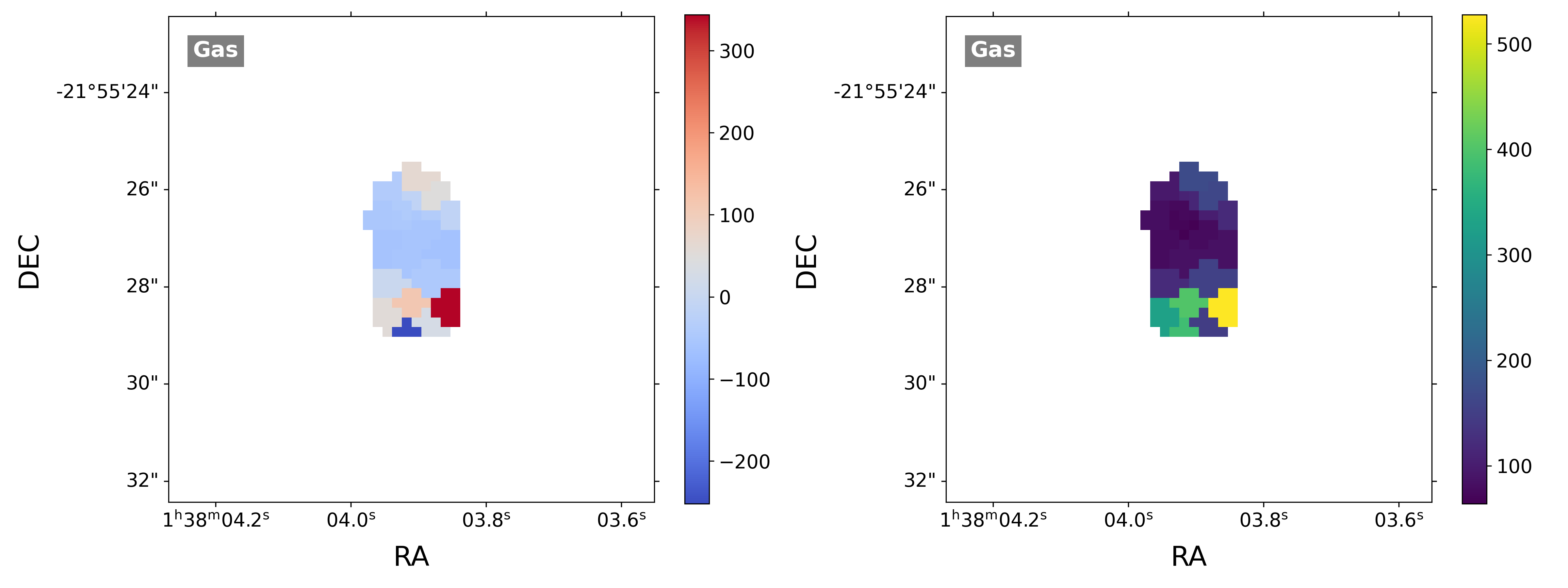}
    \caption{The leftmost panels show the gas velocity offset and gas velocity dispersion in $\mathrm{km s^{-1}}$ spatially resolved per each Voronoi bin in the J2 galaxy tail. The velocity offset is measured with respect to the redshift of the tail ($z=0.3317$). The lower bins in both plots show relatively high values with also larger errors in comparison to bins towards the tip of the tail.}
    \label{fig:J2_kin}
\end{figure*}

\begin{figure*}[t]
    \centering
    \includegraphics[width=\textwidth]{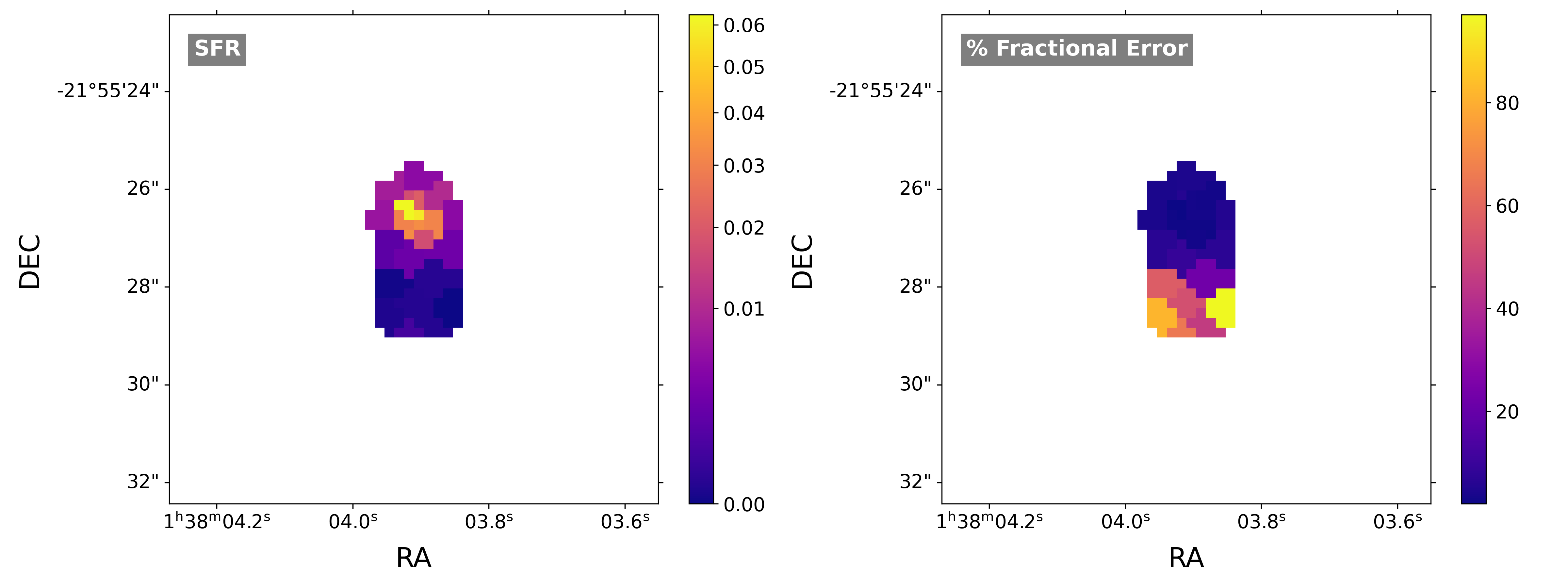}
    \caption{The right panel shows the corrected H$\alpha$-derived SFRs in $\mathrm{M_{\odot}\text{yr}^{-1}kpc^{-2}}$ per each bin with the colorscale shown in square-root stretch. The right plot shows the fractional uncertainties on the star formation rates per each bin.}
    \label{fig:J2_sfr}
\end{figure*}

\subsection{J2}
\label{sec:J2_results}

J2 has a similar structure to J1, with its head spanning 4 kpc and its tail extending 16 kpc. J2 is approximately 15 kpc from the BCG, the closest galaxy in projection to the BCG in our analysis and the most prone to contamination. We follow the same methodology for J1, and choose to utilize the redshift of the head region ($z=0.3308$) for the fitting of continuum-binned regions, and the redshift of the tail region ($z=0.3317$) for the fitting of the H$\alpha+$[NII] emission-binned regions. We again do not use the average redshift of the entire J2 region in our analysis, but report it in Table~\ref{tab:systems}.

The full fit of the entire J2 galaxy is shown in Fig.~\ref{fig:J2_spectra}. No bins are removed from the full fit, and the fit does not change irrespective of the binning method applied. Fig.~\ref{fig:J2_spectra} shows strong Balmer absorption that can be traced mostly to the head of J2, with slight contribution from the lower rightmost region of the tail closest to the head. The strong H$\alpha$ emission can be traced to the remaining tail regions. 

We again utilize the continuum Vorbins to extract stellar kinematics from our \texttt{pPXF} fits. We apply the same statistical cuts on our fits in each Vorbin as described in Section~\ref{sec:spec}, eliminating nearly all regions associated with the tail of J2, except for a couple bins in the lower left corner of the tail with significant absorption. The stellar velocity offsets of each remaining bin relative to the redshift of the head region, as well as the stellar velocity dispersion are shown in Fig.~\ref{fig:J2_head_kin}. We see a slight gradient in the velocity offset across the head, however, there are a couple red bins on the right side that do not follow this gradient. 
Those same bins that have high velocity offsets relative to their neighbors, also have a high velocity dispersion. This is likely due to the head of J2's close proximity to the BCG in this system, and possible contamination of the spectra in this region as a result. Inspection of these individual bins showed an extremely noisy spectrum, though these bins passed our cuts. 
We investigate the potential presence of multiple present stellar populations in Section~\ref{sec:discussion}.

The gas component of J2 is fitted using the Voronoi regions binned for the signal of H$\alpha+$[NII] emission doublet. In this binning scheme, we exclude the head because it did not make the signal cut and fit with the initial redshift of the tail, $z=0.3317$. We choose a normalization range of 5300-5400~\AA, and mask out some larger regions of the observed spectrum from the fit due to noise in the continuum. We utilize an optimal stellar template to account for small absorption features in the spectrum. The gas kinematics for the tail of J2 can be seen in Fig.~\ref{fig:J2_kin}. The bins towards the head, or the base of the tail, have larger errors than those closer to the tip of the tail. The base has much less emission, and therefore less signal to fit from. The fractional uncertainty on the gas velocity dispersion in the lower bins reaches $\sim$20\% of the reported values. Towards the tip of the tail, the fractional uncertainty drops to $\sim$5\% of the reported gas velocity dispersion. 

Both the velocity offset and velocity dispersion show gradients across the tail. While the trend shown is approximately correct, it may not be as drastic as suggested by Fig.~\ref{fig:J2_kin} due to the larger errors in the lower bins. At the tip of the tail where star formation is present, the gas velocity dispersion in that region is $\sim$100$\ \mathrm{km s^{-1}}$. 

The SFRs and the fractional uncertainty per bin are shown in Fig.~\ref{fig:J2_sfr}. There is a clear star-forming knot present at the tip of the tail reaching 0.06 $\mathrm{M_{\odot}\text{yr}^{-1}kpc^{-2}}$
The statistical uncertainty values are $\sim$3\% of the SFR values reported in $\mathrm{M_{\odot}\text{yr}^{-1}kpc^{-2}}$, in the small knot of star formation. In neighboring regions, the uncertainty reaches about $\sim$10\% of the reported SFR values and then increases towards the base of the tail. We left these bins to highlight the structure of star formation; however, the errors are large (reaching nearly $\sim$100\% fractional uncertainty).

\begin{figure*}[htbp]
    \centering
    \includegraphics[width=\textwidth]{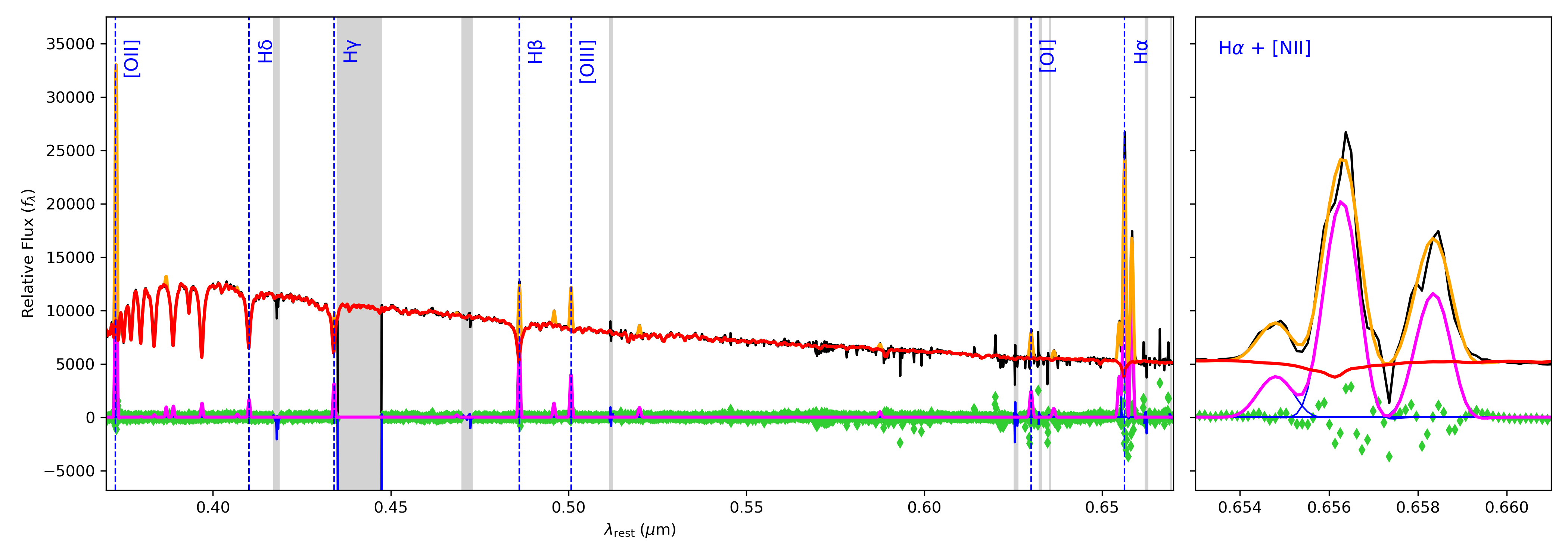}  
    \caption{This figure shows the combined \texttt{pPXF} fit of all of the Voronoi bins of the J3 galaxy. The line colors follow the same scheme set in Fig.~\ref{fig:J1_spectra}. The zoomed-in fit of the H$\alpha+$[NII] emission doublet is shown in the right panel. }
    \label{fig:J3_spectra}
\end{figure*}
\begin{figure*}
    \centering
    \includegraphics[width=\textwidth]{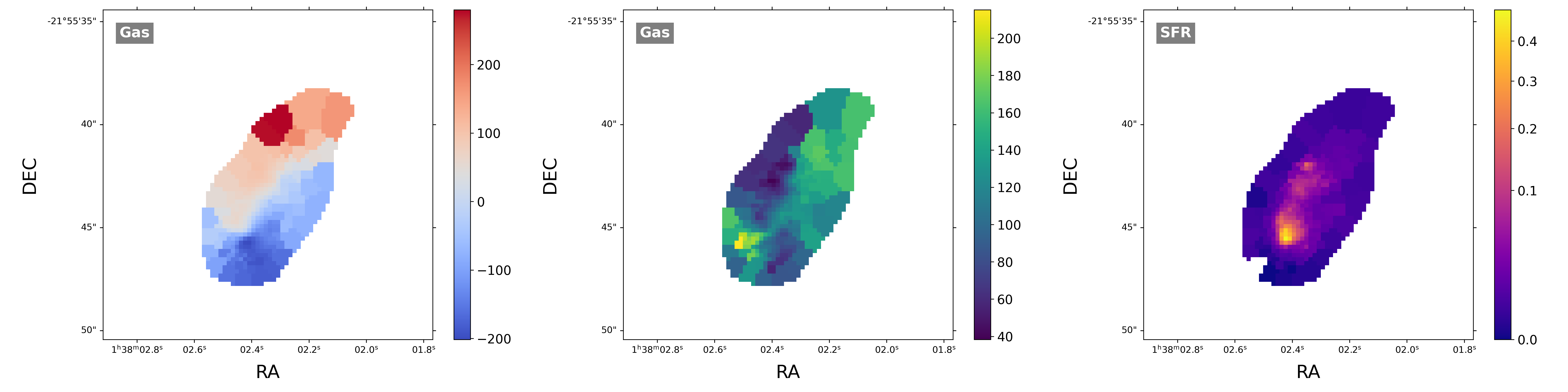}
    \caption{The leftmost panel indicates the gas velocity offset and the middle panel represents the gas velocity dispersion ($\mathrm{km s^{-1}}$) per Voronoi bin from our \texttt{pPXF} fit of each region. The right panel indicates $\text{SFR}_{H_{\alpha}}$ in units of $\mathrm{M_{\odot}\text{yr}^{-1}kpc^{-2}}$, with the color scale shown in square-root stretch. The cutout in the SFR panel in the lower left of J3, is a result of one bin with no signal in the H$\beta$ creating a divide by 0 in the calculation for the star formation in that bin. }
    \label{fig:J3_kin_sfr}
\end{figure*}
\begin{figure*}[htbp]
    \centering
    \includegraphics[width=\textwidth]{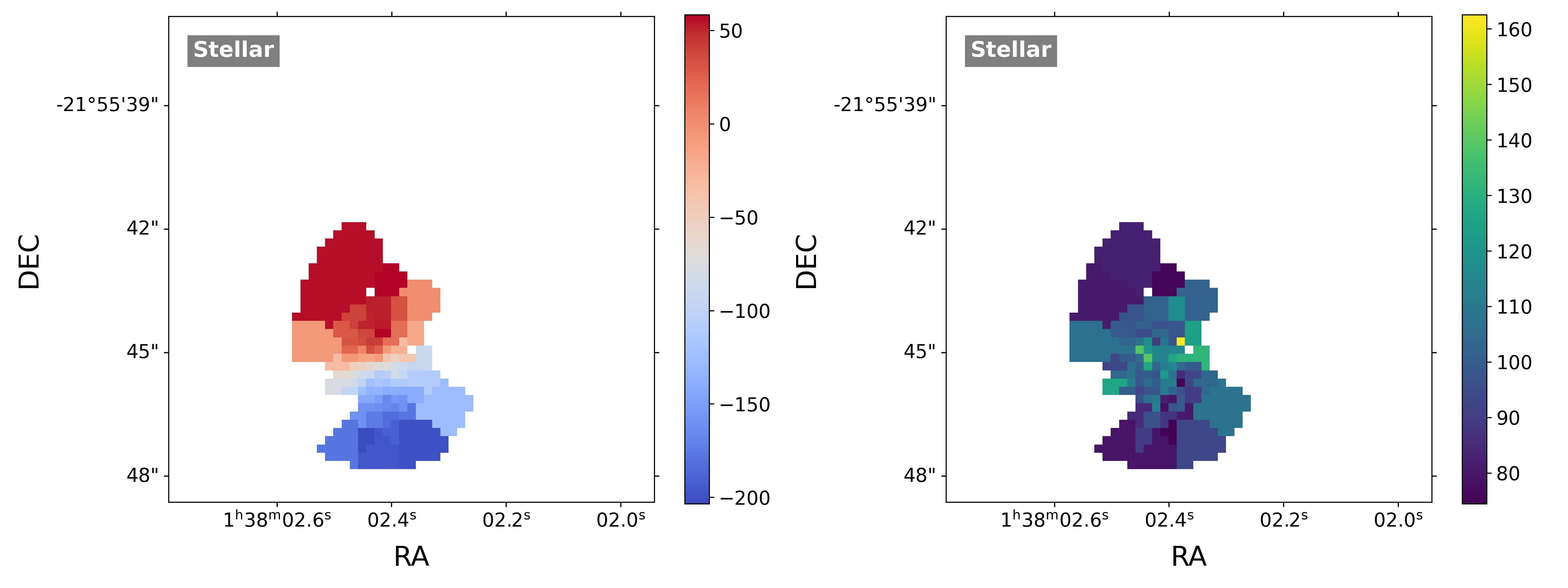}  
    \caption{The line-of-sight stellar velocity offset and dispersion in $\mathrm{km s^{-1}}$ from left to right of Voronoi bins in the J3 galaxy after quality cuts have been applied.}
    \label{fig:J3_stellar_kin}
\end{figure*}

\begin{figure*}[htbp]
    \centering
    \includegraphics[width=\textwidth]{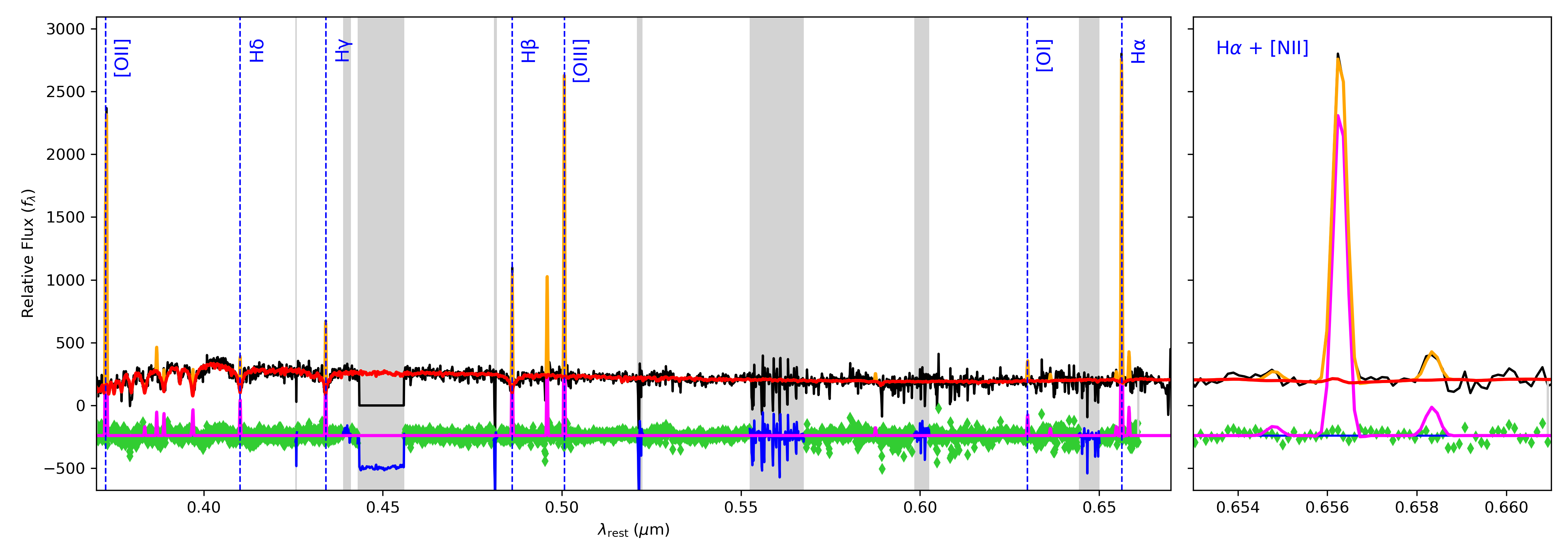}  
    \caption{This figure shows the combined \texttt{pPXF} fit of all of the Voronoi bins in the J4 galaxy. The line color scheme matches that of Fig.~\ref{fig:J1_spectra}. The zoomed-in fit of the H$\alpha+$[NII] emission doublet is shown in the right panel.}
    \label{fig:J4_spectra}
\end{figure*}

\begin{figure*}
    \centering
    \includegraphics[width=\textwidth]{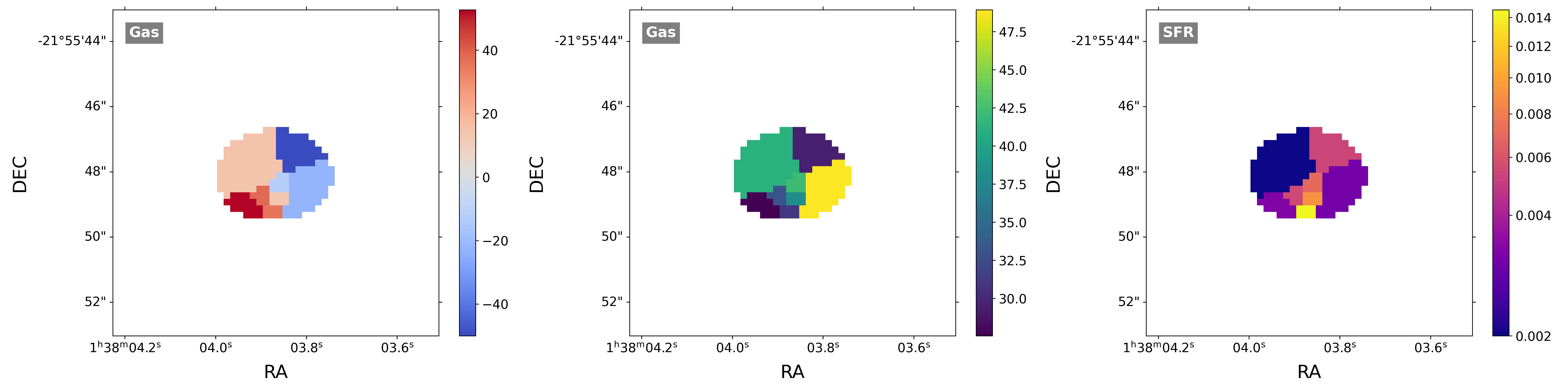}  
    \caption{The leftmost panel indicates the gas velocity offset and the middle panel represents the gas velocity dispersion ($\mathrm{km s^{-1}}$) per Voronoi bin from our \texttt{pPXF} fit of each region in the J4 galaxy. The right panel indicates $\text{SFR}_{H_{\alpha}}$ in units of $\mathrm{M_{\odot}\text{yr}^{-1}kpc^{-2}}$, with the color scale shown in square-root stretch. }
    \label{fig:J4_kin_sfr}
\end{figure*}

\subsection{J3} 
The J3 galaxy lies approximately 110 kpc from the BCG and contains both absorption and emission features. We apply the two binning schemes  for measuring the stellar kinematics and the gas kinematics, respectively as done for the other galaxies. 
Because there is no clear absorption-dominant head or emission-dominant tail, we use the same initial redshift of $z=0.3338$ for our \texttt{pPXF} fits. Both binning scheme fits include both gas and stellar templates to model the observed spectrum. 

The full fit of all Voronoi bins can be seen in Fig.~\ref{fig:J3_spectra}. Some Balmer absorption lines are present in the spectrum and the [OII] and H$\alpha$ emission lines are strong in this galaxy. 

The gas kinematics as well as the SFRs for this galaxy can be found in Fig.~\ref{fig:J3_kin_sfr}. There is a clear gas velocity gradient in the leftmost panel of Fig.~\ref{fig:J3_kin_sfr}, and we conclude this galaxy appears to be in post-pericenter passage, and is now outgoing to the cluster. J3 has the highest star formation rates present in all four galaxies, reaching a value of about 0.49 $\mathrm{M_{\odot}\text{yr}^{-1}kpc^{-2}}$. The statistical uncertainty values on the SFR values are $\sim$5\% of the reported SFR values in $\mathrm{M_{\odot}\text{yr}^{-1}kpc^{-2}}$. The bins towards the bottom edge of J3 had higher errors of $\sim$10-20\% of the reported SFR values, but the errors remained relatively low in the regions of active star formation. We note that one of the bins on the lower left side of J3 in the rightmost panel of Fig.~\ref{fig:J3_kin_sfr} is missing. This is due to a divide by 0 in the estimation of the star formation rate in this bin due to no measured H$\beta$ emission. We also note that the Voronoi bins in the J3 galaxy lie within the composite or LINER regions of our BPT diagram in Fig.~\ref{fig:BPT}, which suggests additional uncertainty on the star formation rates, as not all of the H$\alpha$ emission may be attributable to star formation. We cannot spatially resolve the head from the tail in the J3 galaxy, but the star-forming knot appears to be in the direction of stripping. 

The stellar kinematics for Voronoi bins in J3 with the applied quality cuts are shown in Fig.~\ref{fig:J3_stellar_kin}, where we see a velocity gradient similar to that present in the gas kinematics. The stellar velocity dispersion at the star-forming region is $\sim$100 $\mathrm{km s^{-1}}$ and the gas velocity dispersion in this region appears close to this value as well.

\subsection{J4}
J4 lies approximately 81 kpc from the BCG in projection and was fitted with an initial redshift of $z=0.3087$. We utilize the two binning schemes here for analysis, but only the binning around the H$\alpha +$[NII] doublet yielded multiple Voronoi bins. The continuum binning scheme applied to J4 did not yield more than one bin, at a target S/N per angstrom of 10. This is reflected in Fig.~\ref{fig:Bins}, where the entire region of J4 is highlighted in green. This galaxy was fitted in the wavelength range from 3600-6900~\AA\ due to its lower redshift compared to the previous galaxies. For both fits we used an optimal stellar template and individualized gas templates in each bin. We present the full fit of J4 in Fig.~\ref{fig:J4_spectra}. The spectral fit shows strong emission lines in J4 and some weaker Balmer absorption lines. The gas kinematics in the left two panels of Fig.~\ref{fig:J4_kin_sfr} reveal a slight velocity gradient and may imply rotation. There is some star formation in J4, as shown in the rightmost panel of Fig.~\ref{fig:J4_kin_sfr}, which appears on the side of the galaxy closer to the nearby lensing arc. This is difficult to resolve given the small number of Voronoi bins possible per the target S/N. The star formation reaches a maximum of 0.014 $\mathrm{M_{\odot}\text{yr}^{-1}kpc^{-2}}$ and the statistical uncertainty on the SFRs are $\sim$3\% of the reported SFR values in $\mathrm{M_{\odot}\text{yr}^{-1}kpc^{-2}}$. For our stellar kinematics, we find a stellar velocity dispersion of $35\pm15$ $\mathrm{km s^{-1}}$ for the whole region.

\subsection{BPT Diagram} 
\begin{figure*}[!h]
    \centering
    \includegraphics[width=\textwidth]{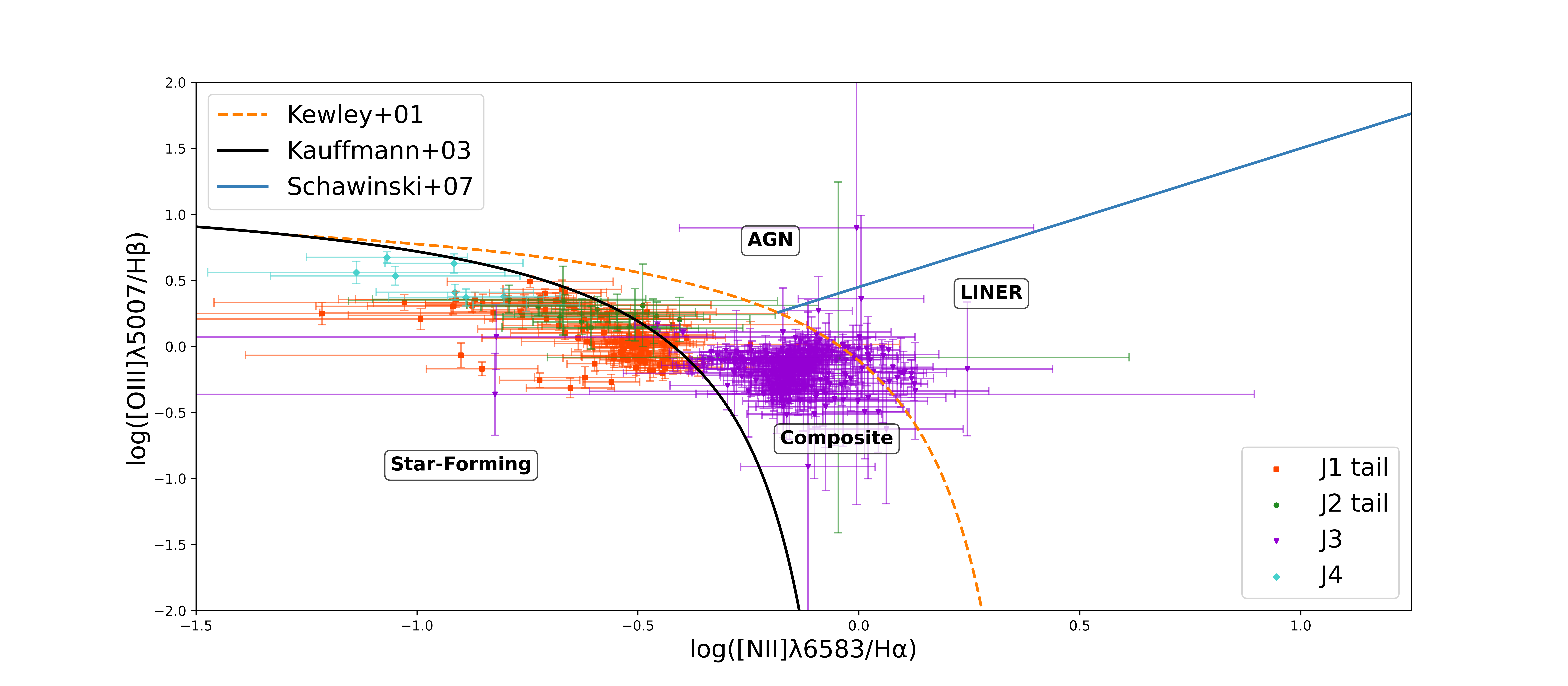}  
    \caption{The [OIII]/H$\beta$ and [NII]/H$\alpha$ BPT line-ratio diagnostics for all Voronoi bins in all jellyfish galaxies. We include the classification lines defined by Kewley, Kauffmann, and Schawinski marked in orange, black, and blue, respectively. We provide statistical error bars for each Voronoi bin. In J1, the largest bin in the tail that lies closest to the head on the right side, as depicted in, was removed from the BPT diagram due to lack of N[II] emission.}
    \label{fig:BPT}
\end{figure*}

We present the BPT diagram ([OIII]$\mathrm{\lambda}$5007/H$\mathrm{\beta}$ versus [NII]$\mathrm{\lambda}$6584/H$\mathrm{\alpha}$) \citep{baldwin1981} for all Voronoi bins containing emission features (in the cases of J1 and J2, these only include the tail bins) in Fig.~\ref{fig:BPT}. We define star-forming, composite, AGN, and LINER regions, separated by the classification lines defined by \cite{kewley2001}, \cite{kauffmann2003}, and \cite{schawinski2007}. Many bins have significant error bars, specifically on the x-axis which is generally a result of low signal in the [NII] region. J2 and J4 fall almost entirely within the purely star-forming region, with a couple J2 bins drifting towards the composite region. J1 also has some bins extending into the composite region. The composite region suggests mixed ionization sources, which may be consistent with the complex environment of ram-pressure stripping where shock heating has the potential to boost emission. 

J3 lies mostly within the composite region with a few bins progressing into the LINER region. To test whether AGN activity drives the extended LINER emission, we examined the spatial distribution of LINER-classified bins across the J3 galaxy. AGN-powered ionization typically produces emission that is strongly concentrated toward the galactic center. However, we find that LINER bins are distributed throughout J3, including regions in the galaxy's outskirts, with no clear radial concentration around the nucleus.  Evidence for shock excitation responsible for extended LINER emission has been shown in galactic gas outflows by \cite{rich2010} in NGC 839, which we consider as a possible explanation for the extended emission we observe for J3 in Fig.~\ref{fig:BPT}. We also point to the gas velocity dispersion values in the middle panel of Fig.~\ref{fig:J3_kin_sfr} as potential evidence for shock ionization, as gas velocity dispersions on the order of $\sim$100 $\mathrm{km s^{-1}}$ (as opposed to lower $\sim$40 $\mathrm{km s^{-1}}$ for pure star formation) have been shown to indicate non-nuclear ionization sources \citep{rich2010, rich2011, epinat2010, ho2014}; however, this alone is not enough to definitively identify shock ionization. Further analysis combining BPT emission line ratios with velocity dispersion and spatial information in 3D space would allow us to better quantify the separation between star formation, AGN, and shock contributions \citep{dagostino2019}. Additionally, other emission line diagnostics like [Fe II]/Pa$\mathrm{\beta}$ could aid in classifying emission from supernova remnants or shocks \citep{alonso-herrero2000}. \cite{Bellhouse2019} used spatially resolved spectroscopy of the jellyfish galaxy JO201 to correlate the ionization state of gas (as traced by BPT line ratios) with physical location, allowing them to identify distinct ionization mechanisms across different galactic regions. A similar spatial-spectroscopic analysis would be valuable for our data if we had sufficient Voronoi bins to resolve this behavior. We conclude that star formation is the dominant ionization source in all galaxies, although shocks resulting from RPS may contribute, especially in the case of J3.

\section{Discussion}
\label{sec:discussion}
In J1, J2, and J3, we observe asymmetric morphologies in the near-infrared and optical HST imaging consistent with the visual structure of jellyfish galaxies described by \cite{ebeling2014}. We observe a characteristic tail of ionized gas, hosting ongoing star formation, that extends away from the primary galactic disk, suggestive of unilateral external forces \citep{poggianti2016}. The tail length as well as the spectral features present in each galaxy allow us to classify the stripping stage for each galaxy as described in \cite{poggianti2025}. 

\begin{figure*}[htbp]
    \centering
    \includegraphics[width=\textwidth]{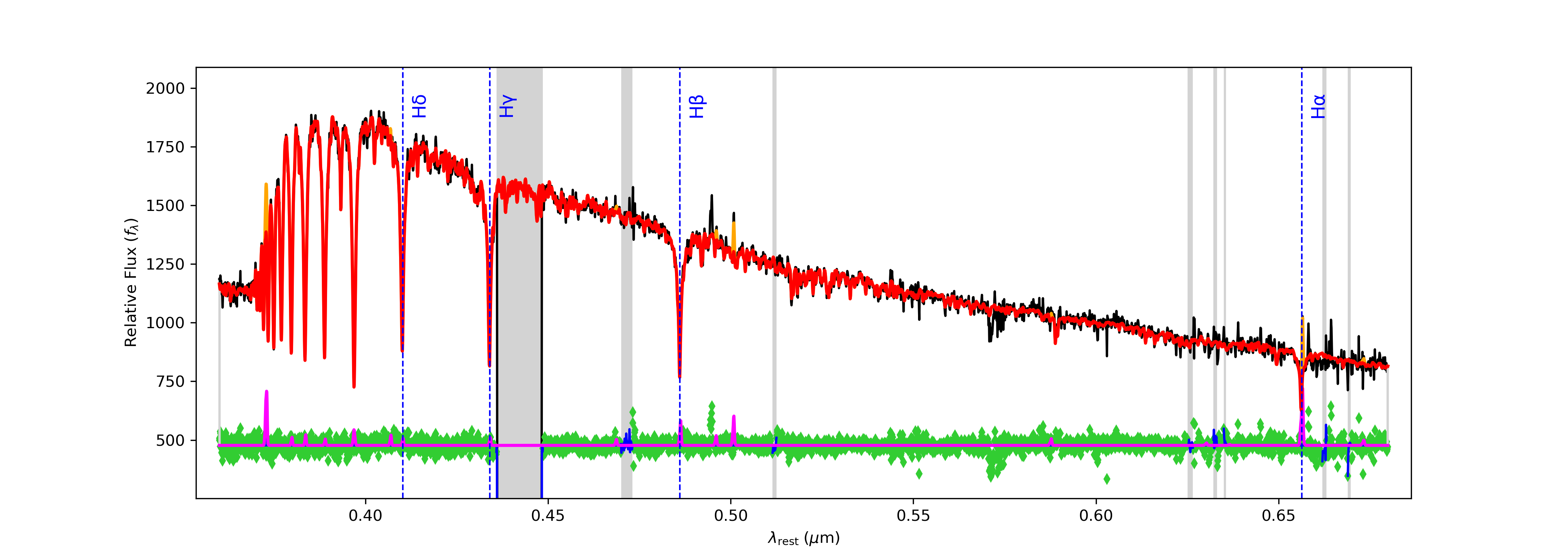}  
    \caption{This figure shows the spectral \texttt{pPXF} fit of just the head region of J1. The line color scheme matches that of Fig.~\ref{fig:J1_spectra}. }
    \label{fig:J1_head_spectra}
\end{figure*}

\begin{figure*}[htbp]
    \centering
    \includegraphics[width=\textwidth]{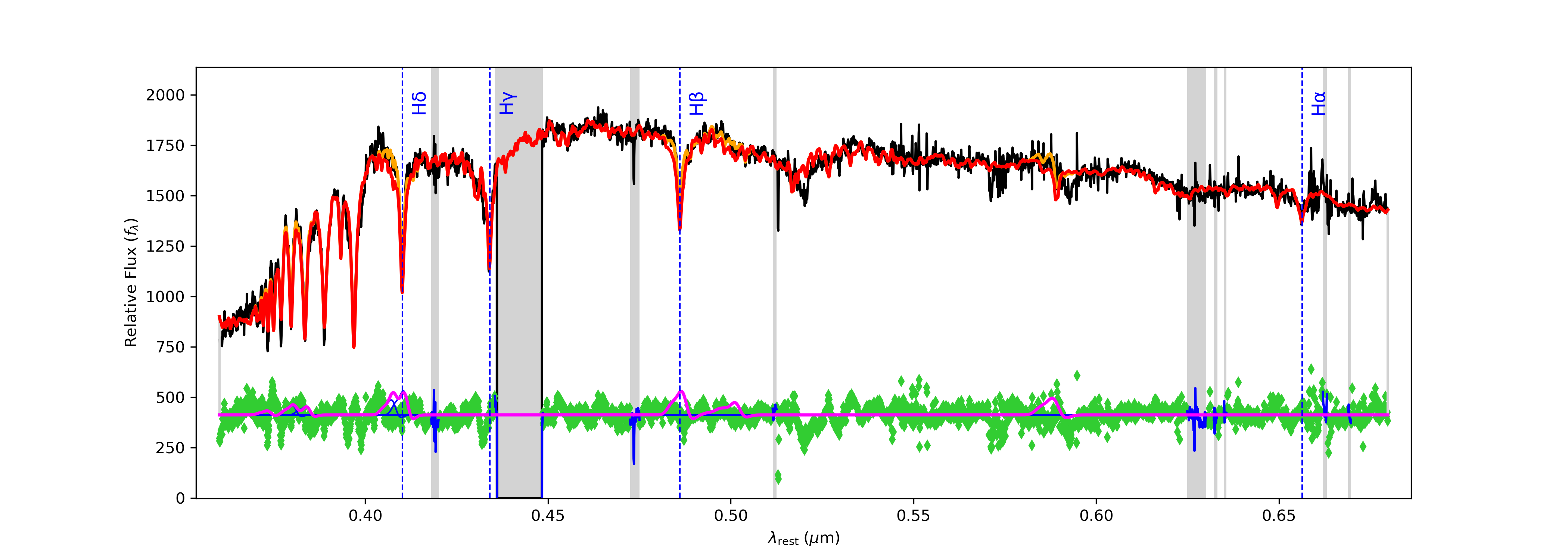}  
    \caption{This figure shows the spectral \texttt{pPXF} fit of just the head region of the J2 galaxy. The line color scheme matches that of Fig.~\ref{fig:J1_spectra}. }
    \label{fig:J2_head_spectra}
\end{figure*}

J1 and J2 have visually distinct heads and elongated tails, evident in the HST imaging, suggesting these two galaxies may be in a late stage of stripping. We provide the fitted spectra for the head regions of J1 and J2 in Fig.~\ref{fig:J1_head_spectra} and Fig.~\ref{fig:J2_head_spectra}, respectively. The spectra of the heads of J1 and J2 are dominated by strong Balmer absorption lines, suggesting A-type stellar populations and point toward quenching within $\sim$1 Gyr \citep{poggianti2004}. We attribute the spectra in the heads of J1 and J2 to post-starburst activity.

In the head of J2, on the redder side of the spectrum in Fig.~\ref{fig:J2_head_spectra} between rest frame $\mathrm{0.50\ \mu m}$-$\mathrm{0.62\ \mu m}$, the fit deviates slightly from the observed spectrum with some absorption features offset from the fit indicating that multiple stellar populations may be present. The head of J2 lies very close in projection to the BCG, and these features are consistent with the BCG redshift. 

We also see some absorption in a few of the bins extending off of the head of J2 on the right side in the continuum binning scheme. The presence of Balmer absorption in two of the tail bins in J2 indicates recent truncation of star formation, consistent with gas stripping and ongoing quenching. The relatively undisturbed stellar component with respect to the galactic head, and lack of detectable absorption features in the rest of the tail suggests that the quenching is primarily driven by ram-pressure stripping rather than tidal interactions \citep{gullieuszik2020}. 
 
The ionized tails of J1 and J2 have strong H$\alpha$ emission, and in the rightmost panel of Fig.~\ref{fig:J1_tail_kinematics}, the J1 tail exhibits a strong in-situ blue star-forming knot in the middle of the tail, with a less prominent knot appearing towards the base of the tail. We also see a star forming knot towards the very tip of the tail in J2 in Fig.~\ref{fig:J2_sfr}. In the classification scheme set by \cite{poggianti2025}, these galaxies might be considered Jtype=2 jellyfish, by the definition that their gas tails are at least as long as the stellar disk, as shown in the HST imaging, although this stage is typically associated with some ionized gas in the head that is not present in the heads of J1 or J2. These two galaxies are clear examples of jellyfish galaxies undergoing ram-pressure stripping. We classify the heads of J1 and J2 as post-starburst galaxies, and attribute the quenching of star formation to ram-pressure stripping, as an obvious ionized tail of stripped gas from these heads is present. Both galaxies also show peculiar velocity gradients across the tail, particularly clear in J1.

The J3 jellyfish galaxy is in a much earlier stage of the stripping process as evident by its visual structure, and combination of absorption and emission features present. The extended tail does cover approximately as much area as the head, although a clear distinction between the head and tail is difficult. We consider this galaxy to be a Jtype=1  or a Jtype=2 jellyfish, where significant stripping is present and the tail seems to cover the length of the head, but we still cannot make a drastic distinction between the head and tail \citep{poggianti2025}. The rightmost panel of Fig.~\ref{fig:J3_kin_sfr} reveals a spatially resolved star-forming knot, although this knot seems to be central in the galaxy in contrast with jellyfish in a later stage which tend to have knots located in their tails. 

J1 and J2 show nearly complete stripping, consistent with efficient RPS at small clustercentric distances in massive clusters where galaxies are stripped most severely \citep{gullieuszik2020}. Despite its larger clustercentric distance, J3 exhibits strong ongoing RPS, likely due to the high cluster mass enabling effective stripping even at greater radii. 

All four galaxies in this analysis exhibit relatively undisturbed stellar kinematics, a prerequisite for recognizing jellyfish galaxies \citep{poggianti2025}. We find in J1, where the galaxy has a distinctive separated head and tail, the velocity dispersion increases as gas travels further downstream the tail. This result is consistent with the velocity dispersion trends observed in ESO137-001, a spiral galaxy undergoing RPS infalling to the Norma Cluster, where velocity dispersions peak around $\sim$100 $\mathrm{km s^{-1}}$ per spaxel around 20 kpc from the galaxy disk \citep{fumagalli2014}. 
Findings from GASP jellyfish galaxy studies, interpret a lower gas velocity dispersion relative to the stellar velocity dispersion in the regions of star formation as evidence that the ionized gas is dynamically cold - creating favorable conditions for in-situ star formation triggered by ram pressure stripping \citep{poggianti2019, gullieuszik2020}. In J3 where we model both stellar and gas kinematics in mostly overlapping regions, we cannot confirm this result but find that our stellar velocity dispersion and gas velocity dispersion in the spatial region of star formation is comparable. 

In the leftmost panel of Fig.~\ref{fig:J1_tail_kinematics}, we see a clear velocity gradient across the tail of J1 extending away from the cluster center. A similar gradient is observed in J3 in Fig.~\ref{fig:J3_kin_sfr}, and we speculate that a similar trend is plausible in the tail of J2, if the uncertainty on the lower gas bins of the tail was lower. This velocity offset gradient is consistent with post-pericenter passage from the BCG on primarily radial orbits \citep{salinas2024, smith2022, jaffe2018}. In J3, the tail appears to extend slightly perpendicular from the cluster center, but further analysis is required to determine the angle between the tail and cluster center \citep{salinas2024}. We observe a slight trend of increasing LOS velocities as gas moves further from the galactic disk, consistent with the RPS classification of galaxies in distant clusters, at approximately z$\sim$0.4, by \citep{moretti2022}. All four galaxies are blue-shifted with respect to the cluster, possibly indicating an infalling group of galaxies. 

It is unclear whether J4 can be characterized as a jellyfish galaxy due to low signal; however, the gas kinematics shown in Fig.~\ref{fig:J4_kin_sfr} reveal J4 appears to be in the foreground of this cluster. The stellar and gas kinematics of J4 as well as the jellyfish galaxies have the potential to inform models of the lensing arcs within MACSJ0138 due to their close proximity to the lensing features.  

J1-3 are strong jellyfish galaxy candidates at z$\sim$0.35, a relatively understudied moderate redshift regime. Detailed characterization of RPS properties at this epoch helps bridge the gap between extensive local studies (z $<$ 0.1) and emerging high-redshift detections at z$\gtrsim$1 \citep[e.g.,][]{boselli2019, roberts2025, xu2025}. These RPS candidates also provide insights into the evolution of star formation efficiency during environmental transformation, as all three are blue-shifted with respect to the cluster. 

\section{Conclusion}
We present an observational study of four galaxies within the strong-lensing MACS J1038.0-2155 cluster at $z=0.336$ utilizing IFU data from MUSE as well as HST imaging in the optical and near-infrared bands. We use adaptive spatial binning based on the signal strength of H$\alpha+$[NII] emission and the continuum region from 6000-6150 \AA\ for each galaxy. To fit the spectra, we use \texttt{pPXF} and extract the stellar and gas kinematics for each galaxy. The main results of our analysis include: 

\begin{enumerate}[label={[\arabic*]}, wide=0pt, leftmargin=*]
    \item We detect strong Balmer absorption lines in the heads of J1 and J2, suggestive of recent quenching. We classify these as post-starburst galaxies. 
    \item We observe gas velocity dispersions in the regions of star formation in all galaxies on the order of $\lesssim 100$ $\mathrm{km s^{-1}}$. 
    We also report stellar velocity dispersions for each galaxy. 
    \item We find prevalent star formation in the center of J3, reaching 0.49 $\mathrm{M_{\odot}\text{yr}^{-1}kpc^{-2}}$ and star formation constrained to the tails of J1 and J2 on the order of 0.10 $\mathrm{M_{\odot}\text{yr}^{-1}kpc^{-2}}$ and 0.010 $\mathrm{M_{\odot}\text{yr}^{-1}kpc^{-2}}$, respectively. We also find star formation activity in J4 on the order of 0.01 $\mathrm{M_{\odot}\text{yr}^{-1}kpc^{-2}}$, but cannot distinguish a characteristic jellyfish structure associated with the star formation.
    \item The velocity offset across the J1, J2, and J3 galaxies indicates they are in post-pericenter passage from the BCG, and occupy approximately radial orbits, consistent with RPS. J4 is in the foreground of the cluster.
    \item We present the BPT diagram for [OIII]$\mathrm{\lambda}$5007/H$\mathrm{\beta}$ versus [NII]$\mathrm{\lambda}$6584/H$\mathrm{\alpha}$ for all Voronoi bins in the tails of J1 and J2 and for J3 and J4. All galaxies fall primarily within purely star-forming or composite regions indicating mostly stellar ionization is present. 
    \item We conclude J1-3 are jellyfish galaxies with tails suggestive of ram-pressure stripping, with J1 and J2 in a later stage of stripping compared to J3. 
\end{enumerate}
In future work, we would like to investigate X-ray data on this system to deepen our understanding of the interaction between the galaxy members and the cluster background in the process of ram-pressure-stripping. 

\begin{acknowledgments}
We thank the Eugene V. Cota-Robles Fellowship for funding this research. We acknowledge Abigail Flowers for productive conversations on the use of \texttt{pPXF} to extract kinematics. 
Software: Numpy \citep{harris2020}, Matplotlib \citep{hunter2007}, Astropy \citep{theastropycollaboration2013, theastropycollaboration2018}, SAO DS9 \citep{joye2017}, Vorbin \citep{cappellari2003}, pPXF \citep{cappellari2004, cappellari2017, cappellari2023}.

\end{acknowledgments}


\bibliographystyle{apsrev4-1}

\nocite{*}

\bibliography{Citations_10_30}

\end{document}